\documentclass[aps,pra,superscriptaddress,twocolumn,10pt]{revtex4-1}
\usepackage{graphicx,amsmath,amssymb}
\usepackage[utf8x]{inputenc}
\usepackage{url}
\usepackage{float}
\usepackage{xcolor}
\usepackage{hyperref}
\usepackage{romannum}

\begin{document}

\title{Intrinsic multipolar contents of nanoresonators for tailored scattering}

\author{Tong Wu}
\affiliation{Univ. Bordeaux, CNRS, Centre de Recherche Paul Pascal, UMR 5031, 33600 Pessac, France}
\affiliation{CNRS, IOGS, Univ. Bordeaux, LP2N, UMR 5298, 33400 Talence, France}

\author{Alexandre Baron}
\affiliation{Univ. Bordeaux, CNRS, Centre de Recherche Paul Pascal, UMR 5031, 33600 Pessac, France}

\author{Philippe Lalanne}
\email{philippe.lalanne@institutoptique.fr}
\affiliation{CNRS, IOGS, Univ. Bordeaux, LP2N, UMR 5298, 33400 Talence, France}

\author{Kevin Vynck}
\email{kevin.vynck@institutoptique.fr}
\affiliation{CNRS, IOGS, Univ. Bordeaux, LP2N, UMR 5298, 33400 Talence, France}

\date{\today}

\begin{abstract}
We introduce a theoretical and computational method to design resonant objects, such as nanoantennas or meta-atoms, exhibiting tailored multipolar responses. In contrast with common approaches that rely on a multipolar analysis of the scattering response of an object upon specific excitations, we propose to engineer the \textit{intrinsic} (i.e., excitation-independent) multipolar content and spectral characteristics of the natural resonances -- or quasinormal modes -- of the object. A rigorous numerical approach for the multipolar decomposition of resonances at complex frequencies is presented, along with an analytical model conveying a direct physical insight into the multipole moments induced in the resonator. Our design strategy is illustrated by designing a subwavelength optical resonator exhibiting a Janus resonance that provides side-dependent coupling to waveguides over the full linewidth of the resonance and on a wide angular range for linearly-polarized incident planewaves. The method applies to all kinds of waves and may open new perspectives for subwavelength-scale manipulation of scattering and emission.
\end{abstract}

\maketitle

Conceiving resonant objects capable of scattering waves along desired directions with a prescribed phase and polarization is pivotal to many applications of wave physics, from acoustic and optical wavefront shaping~\cite{jiang2016convert, ra2017metagratings}, to particle manipulation~\cite{andres2016optical, zaza2019size}, to structural colors engineering~\cite{proust2016all, kristensen2017plasmonic}. At the core of design studies lies the principle that the response of polarizable objects to a driving field can be expanded in terms of radiating multipoles~\cite{bohren2008absorption}. Not only does multipole analysis provide valuable insight into the physical origin of observed scattering features but it spotlights interference conditions between electric and magnetic multipoles that lead to new scattering properties~\cite{yang2017multimode, Picardi2018janus, olmos2019enhanced, shamkhi2019transverse}. An emblematic example is the strong forward scattering that occurs when the electric and magnetic dipole moments induced by an incident field have equal amplitudes and phases~\cite{kerker1983electromagnetic, geffrin2012magnetic} -- creating a so-called ``Huygens source''. The importance of the relative phase between dipole moments was emphasized in a recent study~\cite{Picardi2018janus}, which unveiled a new type of radiating source -- the so-called ``Janus source'' -- providing side-dependent light coupling to waveguides.

The concept of resonance is pivotal to design scattering elements with tailored multipolar behaviors. Resonances, typically identified by peaks of finite linewidth in the spectral response of an object, may be associated to one or several multipoles, depending on the composition, shape and size of the resonator~\cite{wang2006symmetry, verellen2009fano, powell2017interference}. An interference condition may then be reached either from several overlapping resonances with varying multipolar contents~\cite{geffrin2012magnetic, staude2013tailoring, dezert2017isotropic, picardi2019experimental, abdelrahman2019experimental} or from an individual resonance exhibiting alone the desired multipolar content~\cite{asadchy2015broadband}, the latter being more likely to provide features operating over the full linewidth of the resonance. Albeit successful, designs have systematically been achieved until now by a multipolar analysis of the field produced by the resonator \textit{upon excitation by a driving field}~\cite{muhlig2011multipole, evlyukhin2013multipole, bernal2014underpinning}. Because many neighboring resonances are excited at once, even weakly, the dependence of induced moments on the excitation parameters is difficult to apprehend, thereby making the design less intuitive and efficient.

In this Rapid Communication, we introduce a theoretical and computational method for resonator design that is strictly independent of the excitation. Our method, presented here for electromagnetic waves, is based on the concept of quasinormal modes (QNMs), which are the natural resonances of an object, found by solving the source-free Maxwell's equations for the open system. QNM formalisms, which emerged several decades ago~\cite{baum1976singularity,ching1998quasinormal}, have bloomed in recent years, enabling a more insightful and efficient modelling of various problems in photonics and plasmonics~\cite{sauvan2013theory, vial2014resonant, powell2017interference, kewes2018heuristic, yan2018rigorous, franke2019quantization, lalanne2019quasinormal} (for a recent review, see, e.g., Ref.~\cite{lalanne2018light}). Here, we present a rigorous method to compute the \textit{intrinsic} (excitation-independent) multipole moments of individual resonances and derive analytical formulas for the modal decomposition of induced multipole moments at real frequencies. Designs of nanoresonators can thus be achieved largely by analyzing the resonant frequency, linewidth and multipolar content of individual resonances. This possibility is demonstrated by designing a resonator behaving as a Janus source for the scattered field over a resonance linewidth and for a broad range of incident angles.

\begin{figure}
   \centering
   \includegraphics[width=85.77mm]{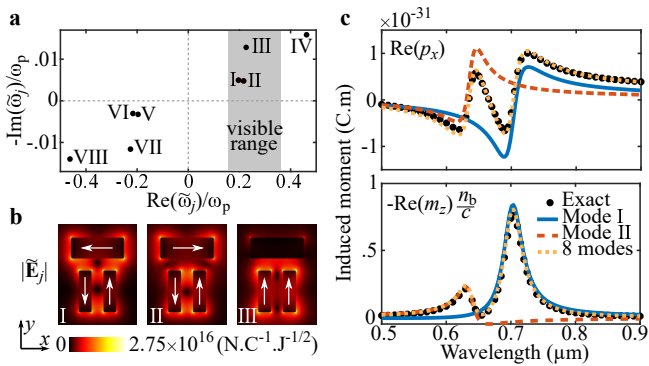}
   \caption{Intrinsic multipolar content of resonances. The plasmonic dolmen is made of silver and composed of an upper rod ($128s  \times 50s \times 20$ nm$^3$) separated by a gap of width $g$ from two lower rods ($30 \times 100 \times 20$ nm$^3$) separated by 30 nm. The silver permittivity is approximated by a single-pole Drude-Lorentz model with $\epsilon_\infty = 1$, $\omega_p = 1.366 \times 10^{16}$ rad.s$^{-1}$, and $\gamma=0.0023 \omega_p$. The dolmen is placed in air ($n_\text{b}=\sqrt{\epsilon_\text{b}}=1$). \textbf{(a)} Complex-frequency plane of the plasmonic dolmen with $g=30$ nm and $s=1$. \textbf{(b)} Spatial maps of $|\tilde{\mathbf{E}}_j|$ of the three QNMs found in the visible range ($j=1$ to $3$). Each QNM behaves as a superposition of electric and magnetic multipoles, as suggested by the current densities (white arrows). The high field enhancement driven by plasmonic effects in the near-field region hides the field divergence due to the complex frequency. \textbf{(c)} Induced dipole moments $\text{Re}[p_x]$ and $-\text{Re}[m_z] n_\text{b}/c$ at real frequencies for a planewave excitation $\mathbf{E}^\text{b}(z) = E_0 \hat{\mathbf{x}} \exp \left[ i \omega n_\text{b} z/c \right]$. The response is dominated by Modes~\Romannum{1} and \Romannum{2} (solid and dashed lines, respectively). Modes~\Romannum{3} and \Romannum{7} are not excited for symmetry reasons. As shown by a comparison with exact real frequency calculations (circles), the induced moments are well predicted with the 8 QNMs represented in panel (a) (dotted lines).}
    \label{fig1}
\end{figure}

\textit{Theory} -- Our formalism is illustrated here without lack of generality with the example of a plasmonic dolmen resonator~\cite{verellen2009fano, bernal2014underpinning}, composed of three nanorods with permittivity $\mathbf{\epsilon}_\text{r}$ placed in a uniform background with permittivity $\epsilon_\text{b}$. The $\exp[-i \omega t]$ convention is used. The QNM formalism unveils the physics underlying light interaction with particles by expanding the field $\mathbf{E}^\text{s}$  scattered by the resonator upon excitation by a driving field $\mathbf{E}^\text{b}$ into a set of natural resonant modes, $\mathbf{E}^\text{s}(\mathbf{r},\omega) = \sum_{j=1}^\infty \alpha_j(\omega) \tilde{\mathbf{E}}_j(\mathbf{r})$. Each QNM is described by a normalized field $\tilde{\mathbf{E}}_j(\mathbf{r})$ and a complex frequency $\tilde{\omega}_j$, and the response to the driving field is fully described via the excitation coefficients $\alpha_j$~\cite{lalanne2018light}. Figures~\ref{fig1}(a)-(b) show the spectral and spatial distributions of the main QNMs of the plasmonic dolmen, computed and normalized using the COMSOL-based solver QNMEig~\cite{yan2018rigorous} of the freeware MAN (Modal Analysis of Nanoresonators)~\cite{note2}, that uses perfectly-matched layers for normalization. Three QNMs are found in the visible range. The current density distributions suggest that Mode~\Romannum{1} behaves as an electric dipole (ED) along $x$ and a magnetic dipole (MD) along $z$, Mode~\Romannum{2} as an ED along $x$ and an electric quadrupole (EQ) in the $xy$-plane, and Mode~\Romannum{3} as an ED along $y$.

The multipolar content of each QNM can be determined quantitatively by expanding its field outside a sphere circumscribing the scatterer in vector spherical wave functions (VSWFs), as
\begin{eqnarray}\label{eq:QNM-VSH-expansion}
\tilde{\mathbf{E}}_j (\mathbf{r}) &=& \tilde{k}_j^2 \sum_{n=1}^\infty \sum_{m=-n}^n E_{nm} \nonumber \\
&\times& \left[ \tilde{a}_{nm,j} \tilde{\mathbf{N}}_{nm,j}^{(3)} (\mathbf{r}) + \tilde{b}_{nm,j} \tilde{\mathbf{M}}_{nm,j}^{(3)} (\mathbf{r}) \right],
\end{eqnarray}
where $\tilde{a}_{nm,j}$ and $\tilde{b}_{nm,j}$ are the electric and magnetic multipole expansion coefficients, $\tilde{\mathbf{N}}_{nm,j}^{(3)}$ and $\tilde{\mathbf{M}}_{nm,j}^{(3)}$ are the outgoing VSWFs and $\tilde{k}_j = \tilde{\omega}_j n_\text{b}/c$ is the complex wavevector of the $j$-th QNM, where $n_\text{b}=\sqrt{\epsilon_\text{b}}$ and $c$ is the light velocity in vacuum. The coefficients $\tilde{a}_{nm,j}$ and $\tilde{b}_{nm,j}$ are obtained by computing the inner product of the QNM field with the VSWFs on the circumscribing sphere surface and the Cartesian multipole moments of each QNM can then be retrieved by matching their far-field expressions with those of the VSWFs in spherical coordinates. This entire procedure is well established for scattered fields at real frequencies~\cite{muhlig2011multipole}. In the present work, the multipolar decomposition, implemented in a dedicated toolbox of MAN~\cite{note2}, is performed \textit{at the QNM complex frequency}. This detail, not mentioned in the one earlier related work~\cite{powell2017interference}, poses no mathematical nor numerical difficulties, but it is indispensable to obtain a mathematically-sound expansion. QNM fields diverge outside the resonator due to the outgoing wave condition $\exp[i \tilde{k}_j r]/r$ with complex $\tilde{k}_j$. As verified numerically in the Supplemental Material (SM)~\cite{SM}, the definition of VSWFs at complex frequencies is necessary to provide unique and stable scattering coefficients as the circumscribing sphere radius increases.

Equation~(\ref{eq:QNM-VSH-expansion}) provides a rigorous description of the multipolar content of a resonance along with a simple way to compute it. However, it relies on an inner product in the near field of the resonator that is difficult to comprehend and does not indicate what remains of the multipolar content at complex frequencies when the resonator is excited by a driving field at real frequencies. To mitigate this lack of physical intuition, we develop an analytical model, valid in the long-wavelength limit, to obtain a decomposition of the induced multipole moments at real frequencies from the intrinsic multipole moments estimated from the QNM fields \textit{inside} the resonator. The derivation, given in detail in SM~\cite{SM}, elaborates on a recent QNM formalism for resonators described by a $N$-pole Drude-Lorentz permittivity as $\epsilon(\omega) = \epsilon_\infty - \epsilon_\infty \sum_{i=1}^N f_i (\omega)$, where $\epsilon_\infty$ is the high-frequency permittivity and $f_i (\omega)$ is the contribution of the $i$-th pole~\cite{yan2018rigorous}. By exploiting the degrees of freedom of auxiliary fields, we arrive to a new QNM expansion for the induced polarization density with new excitation coefficients (compare with Table 1 in Ref.~\cite{lalanne2018light}) that significantly improve the convergence performance on this specific problem compared to alternate truncated QNM expansions~\cite{note1}. This improvement is essential for design purposes, as shown below. In the long-wavelength limit, we obtain QNM expansions for the induced Cartesian multipole moments (here shown only up to the dipole order with $\mathbf{p}$ and $\mathbf{m}$ the electric and magnetic dipole moments, respectively, and for a single pole $N=1$) as
\begin{eqnarray}
\mathbf{p}(\omega) &=& \sum_j \alpha_j(\omega) \nu(\omega,\tilde{\omega}_j) \tilde{\mathbf{p}}_j, \label{eq:QNM-ED-moment} \\
\mathbf{m}(\omega) &=& \sum_j \alpha_j(\omega) \nu(\omega,\tilde{\omega}_j) \frac{\omega}{\tilde{\omega}_j} \tilde{\mathbf{m}}_j, \label{eq:QNM-MD-moment}
\end{eqnarray}
with $\nu(\omega,\tilde{\omega}_j) = \frac{\epsilon(\omega) - \epsilon_\text{b}}{\epsilon(\tilde{\omega}_j) - \epsilon_\text{b}} \frac{\epsilon(\tilde{\omega}_j) - \epsilon_\infty}{\epsilon(\omega) - \epsilon_\infty}$, and where the intrinsic multipole moments are given by

\begin{eqnarray}
\tilde{\mathbf{p}}_j &=& \int_V \epsilon_0 \left[ \epsilon(\tilde{\omega}_j) - \epsilon_\text{b} \right] \tilde{\mathbf{E}}_j(\mathbf{r}) d\mathbf{r},\label{eq:QNM-ED} \\
\tilde{\mathbf{m}}_j &=& -\frac{i \tilde{\omega}_j}{2} \int_V \epsilon_0 \left[ \epsilon(\tilde{\omega}_j) - \epsilon_\text{b} \right] \mathbf{r} \times \tilde{\mathbf{E}}_j(\mathbf{r}) d\mathbf{r}.\label{eq:QNM-MD}
\end{eqnarray}
Equations~(\ref{eq:QNM-ED}) and (\ref{eq:QNM-MD}) are analogous to those known at real frequencies in the long-wavelength limit~\cite{jackson1999classical,terekhov2017multipolar,alaee2018electromagnetic}, indicating indeed that a classical inspection of the QNM field distribution provides a direct visual interpretation of the multipole content. As shown in the SM~\cite{SM}, the moments predicted from Eqs.(\ref{eq:QNM-ED-moment})-(\ref{eq:QNM-MD-moment}) for the plasmonic dolmen differ by less than 5\% compared to those obtained from the VSWF expansion of Eq.~(\ref{eq:QNM-VSH-expansion}), letting us expect that the quasi-static model can be used with confidence for typical plasmonic resonators. Equations~(\ref{eq:QNM-ED-moment}) and (\ref{eq:QNM-MD-moment}) show that the induced multipole moments can be expressed as a linear combination of the intrinsic multipole moments. Note that the $\nu$ coefficient does not appear in previous QNM expansions~\cite{lalanne2018light}. The accuracy of the model is tested in Fig.~\ref{fig1}(c), where we find that the dipole moments induced in the plasmonic dolmen upon planewave excitation are very well explained with only 2 dominant QNMs and quantitatively reproduced with 8 QNMs (out of which 2 are not excited). The possibility to reconstruct the induced moments from only few resonances, which is due to the enhanced convergence rate of our QNM expansion, suggests that \textit{the design of nanoresonators with targeted multipolar response may be performed with a few simulations at complex frequencies without resorting to series of real-frequency simulations}. This will be demonstrated below. One should finally note that electric and magnetic moments differ by a frequency-dependent prefactor $\omega/\tilde{\omega}_j$, implying that electric and magnetic multipole moments, even designed to be perfectly identical at the resonance frequency, cannot be perfectly equal at real frequencies on the resonance linewidth. Nevertheless, since $\text{Im}[\tilde{\omega}_j]<\text{Re}[\tilde{\omega}_j]$ in general ($Q \approx 10$ for most optical resonators), one has $\omega/\tilde{\omega}_j \approx 1$ such that the multipole condition is faithfully reproduced at real frequencies, as will be shown.

\textit{Design} -- Let us now illustrate how the present formalism may be used to design resonators with targeted multipolar responses. Here, we propose to design a resonator that scatters incoming light as a Janus source. Compared to a recent experimental study~\cite{picardi2019experimental}, we aim at a design that is effective on the full linewidth of a resonance for a linearly-polarized incident planewave. The Janus source is composed of electric and magnetic dipoles that have equal amplitudes and are $90^\circ$ out-of-phase~\cite{Picardi2018janus}. As shown above, Mode~\Romannum{1} of the plasmonic dolmen mixes electric and magnetic dipoles that are dephased by about $-\pi/2$. The ratio between the moments is however not equal to 1 and Mode~\Romannum{1} is strongly perturbed by neighboring modes, in particular Mode~\Romannum{2}, leading to a complex lineshape in the induced moments at real frequencies [Fig.~\ref{fig1}(c)].

\begin{figure}
   \centering
   \includegraphics[width=82.89mm]{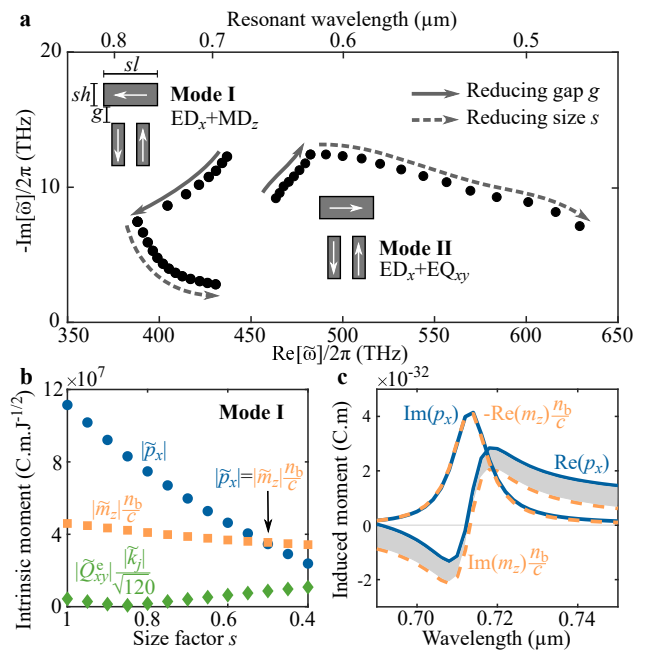}
   \caption{Engineering the spectrum and the multipolar content of the plasmonic dolmen resonances. \textbf{(a)} Trajectories of the two dominant QNMs in the complex frequency plane throughout the design, upon tuning the gap width $g$ from 45 to 10 nm (solid arrows) and the size factor $s$ from 1 to 0.4 (dashed arrows). \textbf{(b)} Mode \Romannum{1} is mostly composed of an electric dipole and a magnetic dipole. For $g=10$ nm and $s = 0.5125$, one obtains $|\tilde{p}_{x,1}| = |\tilde{m}_{z,1}| n_\text{b}/c$ and a phase difference $\Delta \varphi \approx -\pi/2$ (not shown). \textbf{(c)} Induced dipole moments at real frequencies for an $x$-polarized planewave excitation, retrieved from VSWF expansion at real frequencies ($p_x$ in solid lines, $m_z$ in dashed lines). One finds $p_x \approx -i m_z n_\text{b}/c$. The offset (shaded area) in $\text{Re}[p_x]$ is due to the remnants of Mode~\Romannum{2} and other modes outside the visible range, see the SM~\cite{SM}.}
	\label{fig2}
\end{figure}

To achieve a monomode Janus behavior, we tune the dolmen parameters to concurrently reach the interference condition $\tilde{p}_{x,1}=-i\tilde{m}_{z,1}n_\text{b}/c$ for Mode~\Romannum{1} and reduce its spectral overlap with Mode~\Romannum{2}. Note that Mode~\Romannum{3} exhibits a different symmetry compared to Modes~\Romannum{1} and \Romannum{2}, and thus can be ignored by restricting our excitation to $x$-polarized incident planewaves. To spectrally separate Modes~\Romannum{1} and \Romannum{2}, we perform an analysis of the mode trajectories in the complex-frequency plane with varying structural parameters, an approach that is well known in digital filter design~\cite{lathi2005linear} and grating theory~\cite{maystre1982general}. Compared to more recent works on the analysis and design of resonators~\cite{grigoriev2013optimization, romero2016use}, we additionally monitor the multipolar content of each resonance. A full parameter scan in the case of the dolmen would be quite tedious, yet it can be simplified by choosing fewer key parameters. Modes~\Romannum{1} and \Romannum{2} have the same physical origin, that is the coupling between an ED mode in the upper rod and the anti-bonding mode of the bottom dimer. By reducing the distance $g$ between the upper rod and the bottom dimer, and by decreasing the size of the upper rod by a factor $s$, one expects to increase both the level repulsion between the two modes and their quality factors. A decrease of the upper rod size is also expected to decrease the amplitude of the electric dipole moment, which can thus be used to reach the interference condition between intrinsic dipole moments. The impact of these two parameters on the spectral distribution of Modes~\Romannum{1} and \Romannum{2} and the multipolar content of Mode~\Romannum{1} is shown in Fig.~\ref{fig2}(a)-(b) along two specific parameter trajectories. At $g=10$ nm and $s=0.5125$, a condition is reached where the spectral separation between the two modes is much larger than the sum of their linewidths and $|\tilde{p}_{x,1}|=|\tilde{m}_{z,1}| n_\text{b}/c$, while keeping the dephasing very close to $-\pi/2$ (not shown). As a side effect, the EQ of Mode~\Romannum{1} slightly increases, to reach a value that is about 4 times smaller than the ED and MD. However, because the scattering cross-section goes with the absolute square of the moments, the EQ is small enough to be neglected.

To evidence the Janus-like properties of the plasmonic dolmen engineered completely at complex frequencies, we compute the electric and magnetic dipole moments, $p_x$ and $m_z$, induced by illuminating the resonator with an incident $x$-polarized planewave propagating along the $z$-direction. The moments were computed with finite-elements calculations and multipolar decomposition with VSWFs at real frequencies. Results shown in Fig.~\ref{fig2}(c) confirm that $p_x \approx -im_z n_\text{b}/c$ at real frequencies for $\omega \approx \text{Re}[\tilde{\omega}_1]$, thereby validating the design. An offset is observed on $\text{Re}(p_x)$. As shown in the SM~\cite{SM}, this is partly due to Mode~\Romannum{2} which, albeit being spectrally far from Mode~\Romannum{1}, is efficiently excited, but also to other resonances that are farther apart in the spectrum and still contribute weakly with their long Lorentzian tail. As we shall now see, this offset has a weak impact on the performance of the resonator.

\begin{figure}
   \centering
   \includegraphics[width=85.34mm]{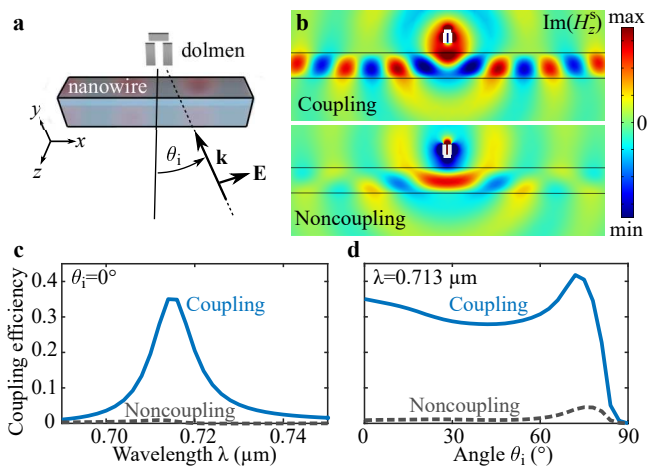}
   \caption{Demonstration of Janus effect in scattering configuration. \textbf{(a)} Sketch of the studied configuration. The dolmen is placed at $117$ nm (border to border) above a dielectric nanowire of index 2.03 with dimensions $270 \times 200$ nm$^2$ along $y$ and $z$ and invariant along $x$, and is illuminated by an $x$-polarized planewave for varying wavelengths and incident angles $\theta_\text{i}$. \textbf{(b)} Spatial maps for the scattered magnetic-field $z$-component at $\theta_\text{i}=0^\circ$ and $\lambda=0.713$ $\mu$m. Depending on the dolmen orientation, light is either efficiently coupled or uncoupled to the fundamental nanowire mode. \textbf{(c)}-\textbf{(d)} Spectral and angular dependence of the coupling efficiencies, evidencing that the Janus effect is effectively implemented for many incidences and frequencies of the driving field.}
    \label{fig3}
\end{figure}

To demonstrate the effectiveness of the present design, we study the side-dependent coupling of the dolmen to the fundamental guided mode of a $\text{Si}_{\text{3}} \text{N}_{\text{4}}$ nanowire (translationally-inviariant along $x$) [Fig.~\ref{fig3}(a)]. As documented in Ref.~\cite{Picardi2018janus}, Janus dipoles offer a unique coupling property that selectively depends on the dipole orientation. Figure~\ref{fig3}(b) shows two maps of the scattered field $H_z^\text{s}=H_z-H_z^\text{b}$ computed for two $\pi$-rotated dolmens at the resonance wavelength $\lambda=0.713$ $\mu$m for an $x$-polarized planewave at normal incidence in the $xz$-plane ($\theta_\text{i} = 0^\circ$). The maps evidence the contrasted coupling behaviors, originally predicted for current sources in Ref.~\cite{Picardi2018janus} and demonstrated here in a scattering configuration. For a more quantitative assessment, we define the coupling efficiency as $C = P^\text{s}/(I_0 \sigma_\text{g})$ where $P^\text{s}$ is the power coupled to the waveguide mode, $I_0$ is the planewave intensity and $\sigma_\text{g}$ is the geometrical cross-section of the smallest circumscribing sphere of the dolmen. $P^\text{s}$ is computed via an overlap integral between the field scattered and the fundamental nanowire mode, see the SM~\cite{SM}. As shown in Figs.~\ref{fig3}(c)-(d), the coupling efficiencies strongly differ, leading to a contrast of about 60 on resonance at normal incidence. Very importantly, this effect is observed over the full linewidth of the resonance, as evidenced by a comparison with Fig~\ref{fig2}(c), and over a wide angular range, up to about $\theta_\text{i} \approx 80^\circ$. In the SM~\cite{SM}, we also show that the designed resonator enables an efficient side-dependent coupling between an $x$-polarized electric dipole source and the dielectric nanowire mode.

\textit{Conclusion} -- By proposing a method for analyzing the multipolar content of individual resonances through an improved QNM formalism~\cite{note2}, we mitigate the limitations encountered in usual resonator designs performed for specific excitations at real frequencies. The proposed method is very effective in terms of computational resources for nanoresonators that are driven by a few resonances only. It also provides a transparent physics, since the multipolar content becomes independent of the driving field and is therefore intrinsically bonded to the natural resonances of the resonator. The force of the method was successfully demonstrated on the example of a Janus behavior in scattering configuration. The design yields a feature that operates on the full linewidth of the resonance and for a wide range of incident angles. To achieve this property, we had to limit the excitation to a specific incident polarization, and yet still observed a small contribution from modes that are distant in the complex-frequency plane. This raises the fundamental question of the potential existence of resonances with a prescribed multipole content that stand alone in the complex plane. Coupled to advanced optimization engines, the present method may allow novel ideas to emerge in quantum optics, nanoscale control of pulses, optical forces and torques, nanoparticle sorting with light, photonic circuits and other devices.

The authors thank Thibault Pichon, Nicolas Mielec and Etienne Hartz for their contributions at different initial stages of the work, and Wei Yan for fruitful discussions. This work has been carried out with financial support from the LabEx AMADEus (Grant No. 10-LABX-0042), and from the ANR projects ``NanoMiX'' (Grant No. ANR-16-CE30-0008), ``Resonance'' (Grant No. ANR-16-CE24-0013) and ``NOMOS'' (Grant No. ANR-18CE24-0026).


\clearpage
\begin{center}
\textbf{\large Supplemental Material for ``Intrinsic multipolar contents of nanoresonators for tailored scattering''}
\end{center}

\setcounter{equation}{0}
\setcounter{figure}{0}
\setcounter{table}{0}
\setcounter{page}{1}
\makeatletter
\renewcommand{\theequation}{S\arabic{equation}}
\renewcommand{\thefigure}{S\arabic{figure}}

\section{Rigorous computation of the multipolar content of quasinormal modes}
\label{secS:QNM-multipoles}

It is well known that the field scattered by a resonator upon excitation (i.e. at real frequency) can be decomposed in terms of vector spherical wave functions at the driving frequency~\cite{mishchenko1996t}. The multipole moments induced in the resonator can then be obtained by matching their far-field behavior with that of radiating multipoles~\cite{muhlig2011multipole}. Here, we straightforwardly generalize this approach to complex frequencies as to compute the multipole moments of radiating QNMs.

We consider the $j$-th QNM of a resonator described by a complex frequency $\tilde{\omega}_j$ and a normalized QNM field $\tilde{\mathbf{E}}_j(\mathbf{r})$. QNMs radiating to free space exhibit a field $\tilde{\mathbf{E}}_j(\mathbf{r})$ that diverges as $\exp[-\text{Im}(\tilde{k}_j)r]/r$ with $\tilde{k}_j=\tilde{\omega}_j n_\text{b}/c$ the complex wavevector of the $j$-th QNM and $\text{Im}(\tilde{k}_j)<0$. It naturally follows that the QNM field outside a sphere circumscribing the resonator can be written as a sum of vector spherical wave functions (VSWFs) defined at the complex frequency $\tilde{\omega}_j$, i.e. with complex wavevector $\tilde{k}_j$, as
\begin{eqnarray}~\label{eqS:multipolar-decomposition}
\tilde{\mathbf{E}}_j (\mathbf{r}) &=& \tilde{k}_j^2 \sum_{n=1}^\infty \sum_{m=-n}^n E_{nm} \nonumber \\
&\times& \left[ \tilde{a}_{nm,j} \tilde{\mathbf{N}}_{nm,j}^{(3)} (\mathbf{r}) + \tilde{b}_{nm,j} \tilde{\mathbf{M}}_{nm,j}^{(3)} (\mathbf{r}) \right],
\end{eqnarray}
where $\tilde{a}_{nm,j}$ and $\tilde{b}_{nm,j}$ are the electric and magnetic multipole expansion coefficients, and $\tilde{\mathbf{N}}_{nm,j}^{(3)}$ and $\tilde{\mathbf{M}}_{nm,j}^{(3)}$ are outgoing VSWFs given by~\cite{mishchenko1996t}
\begin{eqnarray}
\tilde{\mathbf{M}}_{nm,j}^{(3)} (\mathbf{r}) &=& \left[i \pi_{nm}(\cos \theta) \mathbf{e}_\theta - \tau_{nm} (\cos \theta) \mathbf{e}_\varphi \right] \nonumber \\
& \times & h_n^{(1)}(\tilde{k}_j r) \exp(im\varphi), \label{eqS:VSH-M} \\
\tilde{\mathbf{N}}_{nm,j}^{(3)} (\mathbf{r}) &=& n(n+1) P_n^m(\cos \theta) \frac{h_n^{(1)}(\tilde{k}_j r)}{\tilde{k}_j r} \exp(im\varphi) \mathbf{e}_r \nonumber \\
&+& \left[ \tau_{nm} (\cos \theta) \mathbf{e}_\theta + i \pi_{nm}(\cos \theta) \mathbf{e}_\varphi \right] \nonumber \\
& \times & \frac{1}{\tilde{k}_j r} \frac{\textup{d}}{\textup{d}r} \left[ r  h_n^{(1)}(\tilde{k}_j r) \right] \exp(im \varphi)\label{eqS:VSH-N}.
\end{eqnarray}
Here, $h_n^{(1)}$ is the spherical Hankel function of the first kind and $P_n^m$ is the associated Legendre polynomial. $\pi_{nm}$ and $\tau_{nm}$ are defined as
\begin{eqnarray}
\pi_{nm} (\cos \theta) &=& \frac{m}{\sin \theta} P_n^m (\cos \theta), \\
\tau_{nm} (\cos \theta) &=& \frac{\textup{d}}{\textup{d}\theta} P_n^m (\cos \theta).
\end{eqnarray}
The prefactor $E_{nm}$ of Eq.~(\ref{eqS:multipolar-decomposition}) is given by
\begin{equation}
E_{nm} = \frac{1}{2\sqrt{\pi}} i^{n+2m-1} \sqrt{(2n+1)\frac{(n-m)!}{(n+m)!}}.
\end{equation}

The coefficients $\tilde{a}_{nm,j}$ and $\tilde{b}_{nm,j}$ can be obtained via the orthogonality relationship of VSWFs, which is preserved at complex frequencies, leading to
\begin{eqnarray}
\tilde{a}_{nm,j} &=& \frac{\int \tilde{\mathbf{E}}_j (R,\Omega) \cdot \tilde{\mathbf{N}}_{nm,j}^{(3),\star} (R,\Omega) d\Omega}{\tilde{k}_j^2 E_{nm} \int \left| \tilde{\mathbf{N}}_{nm,j}^{(3)} (R,\Omega) \right|^2 d\Omega}, \\
\tilde{b}_{nm,j} &=& \frac{\int \tilde{\mathbf{E}}_j (R,\Omega) \cdot \tilde{\mathbf{M}}_{nm,j}^{(3),\star} (R,\Omega) d\Omega}{\tilde{k}_j^2 E_{nm} \int \left| \tilde{\mathbf{M}}_{nm,j}^{(3)} (R,\Omega) \right|^2 d\Omega}.
\end{eqnarray}
The overlap integrals are performed over the surface of a sphere of a radius $R$ circumscribing the entire particle. Because the VSWFs are defined at the QNM complex frequency, they diverge at large distances from the resonator at the same rate as the QNM fields. This identical behavior makes that $\tilde{a}_{nm,j}$ and $\tilde{b}_{nm,j}$ are independent of $R$.

Once the multipolar coefficients (defined in spherical coordinates) are known, the expressions of the Cartesian moments of a QNM can be retrieved by matching their far-field expressions with those of VSWFs~\cite{bohren2008absorption}. This is again a known derivation at real frequencies~\cite{muhlig2011multipole}. For the lowest order multipole moments, the electric dipole, magnetic dipole and electric quadrupole are respectively given by
\begin{equation}\label{eqS:ED-VSH-decomp}
\tilde{\mathbf{p}}_j = 
\begin{pmatrix}
\tilde{p}_{x,j} \\
\tilde{p}_{y,j} \\
\tilde{p}_{z,j}
\end{pmatrix}
= C_0
\begin{pmatrix}
\tilde{a}_{1-1,j} - \tilde{a}_{11,j}\\
-i (\tilde{a}_{1-1,j} + \tilde{a}_{11,j}) \\ 
-\sqrt{2} \tilde{a}_{10,j}
\end{pmatrix},
\end{equation}
\begin{equation}\label{eqS:MD-VSH-decomp}
\tilde{\mathbf{m}}_j = 
\begin{pmatrix}
\tilde{m}_{x,j} \\
\tilde{m}_{y,j} \\
\tilde{m}_{z,j}
\end{pmatrix}
= \frac{c}{i n_\text{b}}
C_0
\begin{pmatrix}
\tilde{b}_{1-1,j} - \tilde{b}_{11,j}\\
-i (\tilde{b}_{1-1,j} + \tilde{b}_{11,j}) \\ 
-\sqrt{2} \tilde{b}_{10,j}
\end{pmatrix},
\end{equation}
and
\begin{widetext}
\begin{equation}\label{eqS:EQ-VSH-decomp}
\tilde{\mathbf{Q}}_j^e = 
\begin{pmatrix}
\tilde{Q}_{xx,j}^e & \tilde{Q}_{xy,j}^e & \tilde{Q}_{xz,j}^e \\
\tilde{Q}_{yx,j}^e & \tilde{Q}_{yy,j}^e & \tilde{Q}_{yz,j}^e \\
\tilde{Q}_{zx,j}^e & \tilde{Q}_{zy,j}^e & \tilde{Q}_{zz,j}^e
\end{pmatrix}
= D_0
\begin{pmatrix}
i (\tilde{a}_{22,j}+\tilde{a}_{2-2,j}) - \frac{i \sqrt{6}}{3} \tilde{a}_{20,j} & (\tilde{a}_{2-2,j} - \tilde{a}_{22,j}) & i(\tilde{a}_{21,j}-\tilde{a}_{2-1,j})\\
(\tilde{a}_{2-2,j} - \tilde{a}_{22,j}) & - i (\tilde{a}_{22,j}+\tilde{a}_{2-2,j}) - \frac{i \sqrt{6}}{3} \tilde{a}_{20,j} & -(\tilde{a}_{21,j}+\tilde{a}_{2-1,j}) \\
i(\tilde{a}_{21,j}-\tilde{a}_{2-1,j}) & -(\tilde{a}_{21,j}+\tilde{a}_{2-1,j}) & \frac{2\sqrt{6}i}{3} \tilde{a}_{20,j}
\end{pmatrix}
\end{equation}
\end{widetext}
with $C_0=\sqrt{6 \pi \epsilon_0} n_\text{b}^2 i/(c\sqrt{\mu_0} \tilde{k}_j)$ and $D_0=6 \sqrt{30 \pi \epsilon_0} n_\text{b}^2 /(ic\sqrt{\mu_0} \tilde{k}_j^2)$. $\epsilon_0$ and $\mu_0$ are the vacuum permittivity and permeability, respectively. Here, the formulas are different from those in Ref.~\cite{muhlig2011multipole} due to the associated Legendre polynomials used in Eqs.~(\ref{eqS:VSH-N})-(\ref{eqS:VSH-N}) as
\begin{equation}
P_n^m(x) = (-1)^m (1-x^2)^{m/2} \frac{\textup{d}^m}{\textup{d}x^m} P_n(x),
\end{equation}
where $P_n$ are the unassociated Legendre polynomials.

\begin{figure}
   \centering
   \includegraphics[width=77.6762mm]{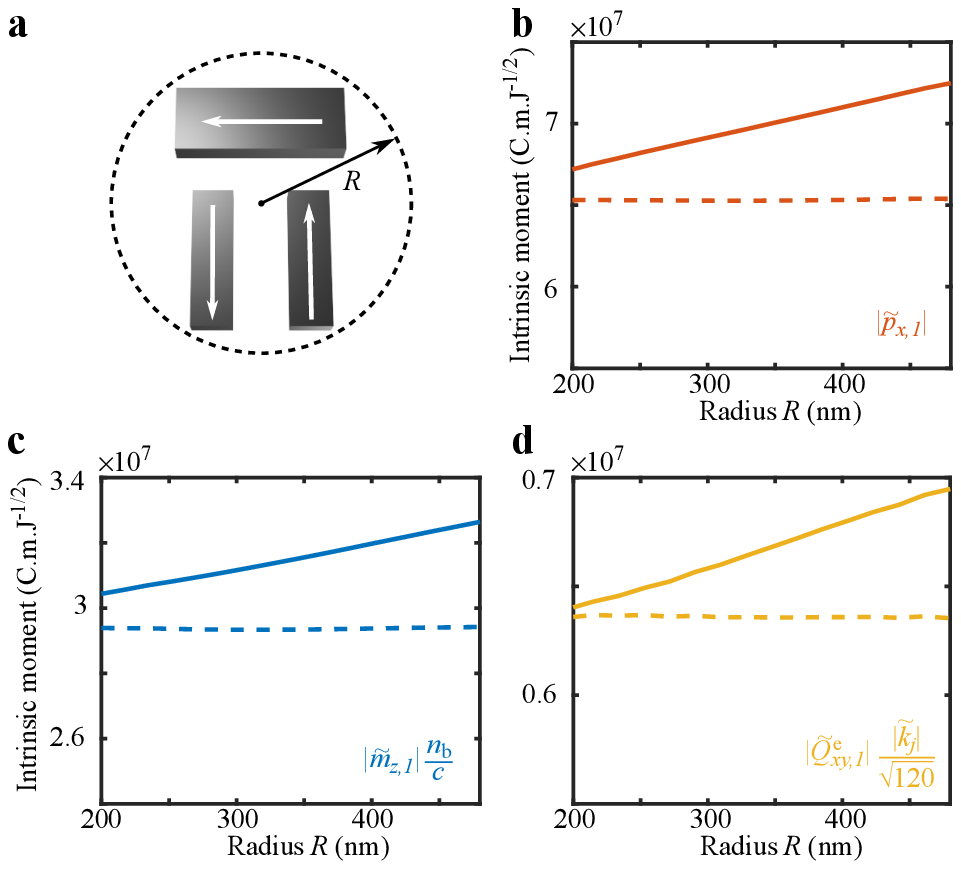}
   \caption{Dependence of multipole moments of a QNM on radius $R$ of the circumscribing sphere. The investigated resonator is the plasmonic dolmen, with the same geometric and material parameters as in Fig. 1 of the main text, with $g=30$ nm and $s=1$. The figures show the absolute values of the electric dipole moment $|\tilde{p}_{x,j}|$, magnetic dipole moment $|\tilde{m}_{z,j}|$ and electric quadrupole moment $|\tilde{Q}^\text{e}_{xy,j}|$ for mode 1 ($j=1$). Mode~\Romannum{1} is found at the complex frequency $\tilde{\omega}_j/2 \pi=  4.269 \times 10^{14} - i1.083 \times 10^{13}$ Hz. The solid lines and dashed lines show the moments computed by performing the multipolar decomposition with VSWFs with real wavevectors $k = \text{Re}[\tilde{k}_j]$ and with complex wavevectors $\tilde{k}_j$, respectively. Only the latter provides quantitatively correct results since the moments do not depend on $R$.}
       \label{figS1}
\end{figure}

To show the importance of using a complex wavevector in the multipolar decomposition, we take the plasmonic dolmen considered in this work and compute the Cartesian multipole moments of Mode~\Romannum{1} as a function of the radius $R$ of the sphere on which the VSWF expansion is made. Results are shown in Fig.~\ref{figS1} for the main three multipole components, i.e. $\tilde{p}_x$, $\tilde{m}_z$ and $\tilde{Q}^\text{e}_{xy}$. We observe that, as expected, the computed moment amplitudes are independent of $R$ when the VSWFs are defined with a complex wavevector $\tilde{k}_j$. On the other hand, if we define the VSWFs as functions of the real part of the mode frequency $\text{Re}\left[\tilde{\omega}_j \right]$, the moment amplitudes diverge. While the ratio between the various moments remains constant for a specific mode~\cite{powell2017interference}, their absolute amplitudes have no physical meaning. This implies that they could not be used to predict the actual response of a resonator upon excitation by a driving field, as demonstrated in this paper.

\section{Quasinormal-mode formalism for an intuitive analysis of multipolar behavior of small electromagnetic resonators}
\label{secS:multipole-QNM-theory}

\subsection{Theory}

In this part, we provide a detailed derivation of the expressions reported in the main text, defining Cartesian multipole moments of quasinormal modes (QNMs) and how they are related to the multipole moments induced in the resonator at real frequencies. The polarization density in a resonator is defined as $\mathbf{P}(\mathbf{r},\omega) = \epsilon_0 [\epsilon(\mathbf{r},\omega) - \epsilon_\text{b}] \mathbf{E}(\mathbf{r},\omega)$ where $\epsilon(\mathbf{r},\omega)$ is the relative permittivity of the resonator, $\epsilon_\text{b}$ is the permittivity of the uniform background and $\mathbf{E}$ is the total field. At optical frequencies, many material permittivities can be described by a $N$-pole Drude-Lorentz relationship~\cite{ashcroft1976solid}. Assuming that the permittivity is constant over the resonator volume $V$, and in line with Ref.~\cite{yan2018rigorous}, we define
\begin{equation}
\epsilon(\omega) = \epsilon_\infty - \epsilon_\infty \sum_{i=1}^N f_i (\omega),
\end{equation}
with $\epsilon_\infty$ the high-frequency permittivity and $f_i (\omega) = \omega_{\text{p},i}^2/ \left(\omega^2 - \omega_{r,i}^2 + i \omega \gamma_i \right)$, where $\omega_{\text{p},i}$ is the plasma frequencies, $\omega_{r,i}$ the resonant frequency and $\gamma_i$ the damping coefficient of the $i$-th pole. The polarization density associated to the $i$-th pole is thus
\begin{equation}
\mathbf{P}_i(\mathbf{r},\omega) = - \epsilon_0 \epsilon_\infty f_i (\omega) \mathbf{E}(\mathbf{r},\omega),
\end{equation}
such that
\begin{equation}\label{eqS:P_realomega}
\mathbf{P}^\text{r}(\mathbf{r},\omega) = \sum_{i=1}^N \mathbf{P}_i(\mathbf{r},\omega) = \epsilon_0 \left[ \epsilon(\omega) - \epsilon_\infty \right] \mathbf{E}(\mathbf{r},\omega),
\end{equation}
and $\mathbf{P} = \mathbf{P}^\text{r} + \epsilon_0 [ \epsilon_\infty - \epsilon_\text{b}]\mathbf{E}$. Following the rigorous modal analysis of resonators described in Ref.~\cite{yan2018rigorous} and considering a single-pole permittivity ($N=1$) to simplify notations, we expand the auxiliary field $\mathbf{P}_{i=1} \equiv \mathbf{P}^\text{r}$ in terms of QNMs as
\begin{equation}\label{eqS:P_expansion}
\mathbf{P}^\text{r}(\mathbf{r},\omega) = \sum_j \alpha_j(\omega) \tilde{\mathbf{P}}_j(\mathbf{r}),
\end{equation}
where $\alpha_j(\omega)$ is the excitation coefficient of the $j$-th QNM, given by~\cite{yan2018rigorous}
\begin{eqnarray}
\alpha_j(\omega) &=& \frac{\tilde{\omega}_j}{\tilde{\omega}_j - \omega} \int_V \left[ \epsilon(\tilde{\omega}_j)-\epsilon_\text{b} \right] \tilde{\mathbf{E}}_j (\mathbf{r}) \cdot \mathbf{E}^\text{b} (\mathbf{r},\omega) d\mathbf{r} \nonumber \\
&+& \int_V \left[ \epsilon_\text{b}-\epsilon_\infty \right] \tilde{\mathbf{E}}_j (\mathbf{r}) \cdot \mathbf{E}^\text{b} (\mathbf{r},\omega) d\mathbf{r},
\end{eqnarray}
and $\tilde{\mathbf{P}}_{j}$ is the normalized auxiliary field of the $j$-th QNM, defined as
\begin{equation}\label{eqS:P_j}
\tilde{\mathbf{P}}_{j}(\mathbf{r}) = \epsilon_0 \left[ \epsilon(\tilde{\omega}_j) - \epsilon_\infty \right] \tilde{\mathbf{E}}_j(\mathbf{r}).
\end{equation}
Inserting Eq.~(\ref{eqS:P_j}) into Eq.~(\ref{eqS:P_expansion}) and comparing the resulting expression with Eq.~(\ref{eqS:P_realomega}) leads to a QNM expansion for the internal field
\begin{equation}\label{eqS:internal-field-QNM}
\mathbf{E}(\mathbf{r},\omega) = \sum_j \alpha_j(\omega) \frac{\epsilon(\tilde{\omega}_j) - \epsilon_\infty}{\epsilon(\omega) - \epsilon_\infty} \tilde{\mathbf{E}}_j(\mathbf{r}).
\end{equation}
which, using Eq.~(\ref{eqS:P_j}), leads to a QNM expansion for the polarization density
\begin{equation}\label{eqS:polarization-density-QNM}
\mathbf{P}(\mathbf{r},\omega) = \sum_j \alpha_j(\omega) \frac{\epsilon(\omega) - \epsilon_\text{b}}{\epsilon(\omega) - \epsilon_\infty} \tilde{\mathbf{P}}_j(\mathbf{r}).
\end{equation}
Equations~(\ref{eqS:internal-field-QNM}) and (\ref{eqS:polarization-density-QNM}) are completely original to our knowledge. We have observed on the specific case of the plasmonic dolmen that they lead to good convergence of the reconstructed internal field and polarization density compared to other expansions. Systematic tests and comparisons will be provided in a future work.

To reach physically-intuitive relations for the multipole moments, we take the long-wavelength approximation and assume that the toroidal multipole can be ignored~\cite{terekhov2017multipolar,alaee2018electromagnetic}. At real frequencies, the electric and magnetic dipole moments and electric quadrupole moment induced in the resonator are given by~\cite{jackson1999classical}
\begin{eqnarray}
\mathbf{p}(\omega) &\approx& \int_V \mathbf{P}(\mathbf{r},\omega) d\mathbf{r},\label{eqS:ED-total-field} \\
\mathbf{m}(\omega) &\approx& - \frac{i\omega}{2} \int_V \mathbf{r} \times \mathbf{P}(\mathbf{r},\omega) d\mathbf{r},\label{eqS:MD-total-field} \\
\mathbf{Q}^\text{e}(\omega) &\approx& 3 \int_V \bigg[ \mathbf{r}\mathbf{P}(\mathbf{r},\omega) + \mathbf{P}(\mathbf{r},\omega)\mathbf{r} \nonumber \\
&-& \frac{2}{3} \mathbf{r} \cdot \mathbf{P} (\mathbf{r},\omega) \mathbf{I} \bigg] d\mathbf{r}, \label{eqS:EQ-total-field}
\end{eqnarray}
where $\mathbf{a}\mathbf{b}$ is the tensor product between $\mathbf{a}$ and $\mathbf{b}$, and $\mathbf{I}$ is the unit dyadic tensor. Inserting Eq.~(\ref{eqS:polarization-density-QNM}) with Eq.~(\ref{eqS:P_j}) into Eqs.~(\ref{eqS:ED-total-field})-(\ref{eqS:EQ-total-field}), and introducing the Cartesian multipole moments of QNMs as
\begin{eqnarray}
\tilde{\mathbf{p}}_j &\approx& \int_V \epsilon_0 \left[ \epsilon(\tilde{\omega}_j) - \epsilon_\text{b} \right] \tilde{\mathbf{E}}_j(\mathbf{r}) d\mathbf{r},\label{eqS:QMN-ED} \\
\tilde{\mathbf{m}}_j &\approx& -\frac{i \tilde{\omega}_j}{2} \int_V \epsilon_0 \left[ \epsilon(\tilde{\omega}_j) - \epsilon_\text{b} \right] \mathbf{r} \times \tilde{\mathbf{E}}_j(\mathbf{r}) d\mathbf{r},\label{eqS:QMN-MD} \\
\tilde{\mathbf{Q}}^\text{e}_j &\approx&  3 \int_V \epsilon_0 \left[ \epsilon(\tilde{\omega}_j) - \epsilon_\text{b} \right] \nonumber \\
& \times & \left[ \mathbf{r} \tilde{\mathbf{E}}_j(\mathbf{r}) + \tilde{\mathbf{E}}_j(\mathbf{r})\mathbf{r} - \frac{2}{3} \mathbf{r} \cdot \tilde{\mathbf{E}}_j(\mathbf{r}) \mathbf{I} \right] d\mathbf{r}\label{eqS:QMN-EQ},
\end{eqnarray}
we immediately reach
\begin{eqnarray}
\mathbf{p}(\omega) &\approx& \sum_j \alpha_j(\omega) \nu(\omega,\tilde{\omega}_j) \tilde{\mathbf{p}}_j,   \label{eqS:QMN-ED-moment} \\
\mathbf{m}(\omega) &\approx& \sum_j \alpha_j(\omega) \nu(\omega,\tilde{\omega}_j) \frac{\omega}{\tilde{\omega}_j}\tilde{\mathbf{m}}_j, \label{eqS:QMN-MD-moment} \\
\mathbf{Q}^\text{e} (\omega) &\approx& \sum_j \alpha_j(\omega) \nu(\omega,\tilde{\omega}_j) \tilde{\mathbf{Q}}^\text{e}_j, \label{eqS:QMN-EQ-moment},
\end{eqnarray}
with $\nu(\omega,\tilde{\omega}_j) = \frac{\epsilon(\omega) - \epsilon_\text{b}}{\epsilon(\tilde{\omega}_j) - \epsilon_\text{b}} \frac{\epsilon(\tilde{\omega}_j) - \epsilon_\infty}{\epsilon(\omega) - \epsilon_\infty}$, which are the expressions given in the main text. For systems with $\epsilon_\text{b} = \epsilon_\infty$, the moments further simplify to
\begin{eqnarray}
\mathbf{p}(\omega) &\approx& \sum_j \alpha_j(\omega) \tilde{\mathbf{p}}_j, \label{eqS:ED-induced-QNM} \\
\mathbf{m}(\omega) &\approx& \sum_j \alpha_j(\omega) \frac{\omega}{\tilde{\omega}_j} \tilde{\mathbf{m}}_j, \label{eqS:MD-induced-QNM} \\
\mathbf{Q}^\text{e} (\omega) &\approx& \sum_j \alpha_j(\omega) \tilde{\mathbf{Q}}^\text{e}_j. \label{eqS:EQ-induced-QNM}
\end{eqnarray}

\subsection{Comparison of QNM multipole moments with exact predictions}

\begin{figure}
   \centering
   \includegraphics[width=44.2324mm]{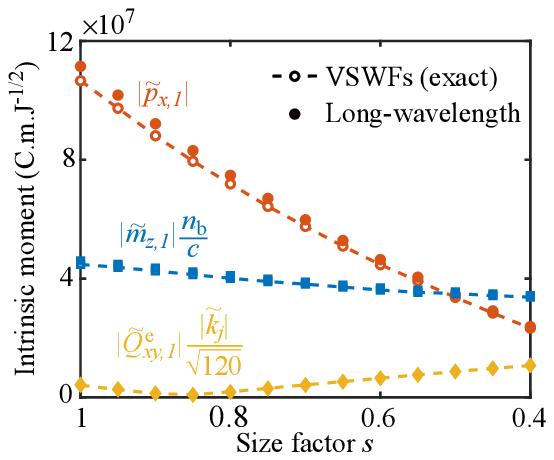}
   \caption{Comparison of the two approaches to compute the multipole moments of Mode \Romannum{1} of the plasmonic dolmen studied in the main text as a function of the size factor $s$. The filled markers have been obtained from the internal QNM field using Eqs.~(\ref{eqS:QMN-ED})-(\ref{eqS:QMN-EQ}) -- corresponding to the results presented in Fig.~2(b) of the main text -- and the dashed lines with empty markers from the VSWF expansion of the external QNM field using Eqs.~(\ref{eqS:ED-VSH-decomp})-(\ref{eqS:EQ-VSH-decomp}). A very good agreement is observed, especially for small values of the size factor $s$.}
       \label{figS2}
\end{figure}

Let us now compare the predictions of the multipole moments as obtained from the internal field using Eqs.~(\ref{eqS:QMN-ED})-(\ref{eqS:QMN-EQ}) and from the VSWF expansion of the QNM field outside the resonator using Eqs.~(\ref{eqS:ED-VSH-decomp})-(\ref{eqS:EQ-VSH-decomp}). Results are shown in Fig.~\ref{figS2} for Mode~\Romannum{1} of the plasmonic dolmen as a function of the size factor $s$ of the top rod, thereby reproducing Fig.~2(b) in the main text. A very good agreement is observed between the two methods, especially for small values of $s$. The increasing error for the electric dipole moment with increasing size factors is due to the progressive failure of the long wavelength approximation used in Eqs.~(\ref{eqS:QMN-ED})-(\ref{eqS:QMN-EQ}). Thus, we conclude (i) that both approaches to compute the multipole moments of QNMs are consistent with each other and (ii) that the long-wavelength approximation is a good approximation for the study of the plasmonic dolmen. Let us also note that the computational load required to compute the intrinsic multipole moments are not significantly different between the two methods. The clear advantage of the long-wavelength formula, Eqs.~(\ref{eqS:QMN-ED})-(\ref{eqS:QMN-EQ}), lies in their explicit dependence on the QNM fields, which helps apprehending the physical origin of their intrinsic multipolar content.

\subsection{Recovery of the induced moments at real frequencies for the engineered dolmen}

\begin{figure}
   \centering
      \includegraphics[width=81.2325mm]{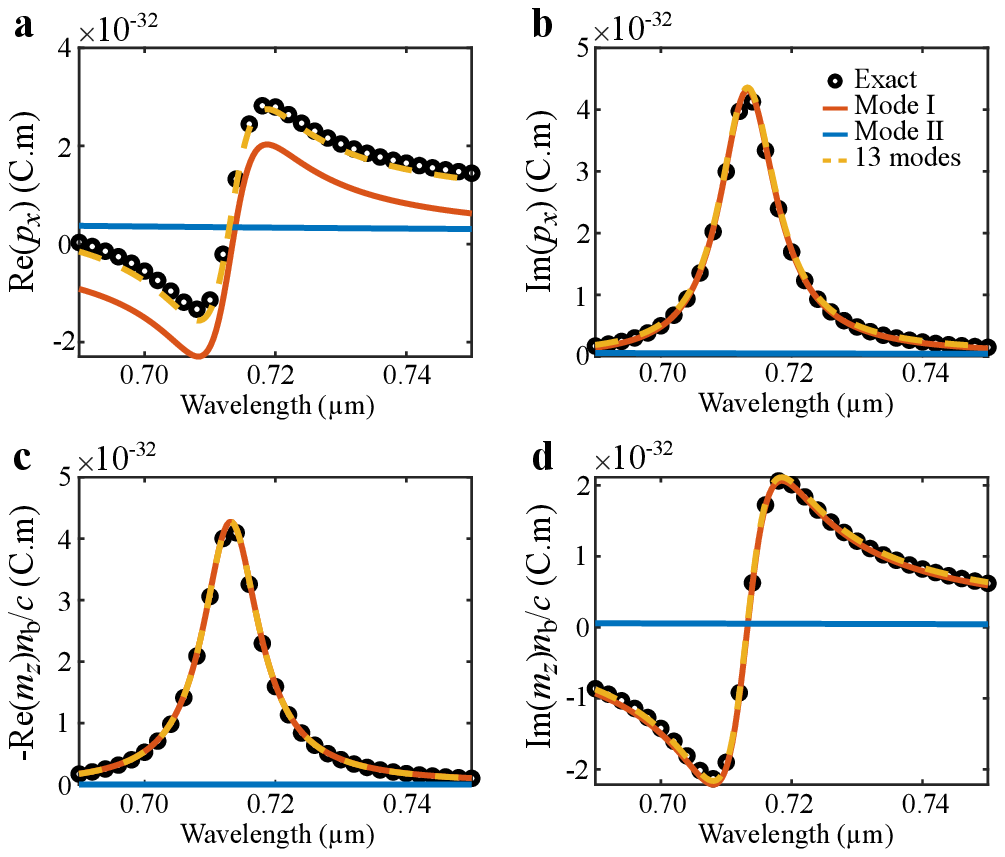}
   \caption{Reconstruction of the induced multipole moments of the plasmonic dolmen with QNMs. \textbf{(a)-(d)} Spectra of the real and imaginary parts of the electric dipole $p_x$ (a)-(b) and magnetic dipole moment $m_z$ (c)-(d) induced in the resonator upon planewave excitation. Mode~\Romannum{1} clearly brings the largest contribution to the induced dipole moments in this frequency range. Mode~\Romannum{2} has a small yet non-negligible contribution on $\text{Re}[p_x]$. The exact results (empty markers) are predicted with good accuracy by summing over 13 QNMs.}
       \label{figS3}
\end{figure}

In Fig. 1(c) of the main text, we show that the induced moments at real frequencies can be reconstructed from the intrinsic moments at complex frequencies in a non-engineered dolmen. Here, we perform this similar analysis for the engineered dolmen, showing in particular that the offset in $\text{Re}[p_x]$ is indeed due to other modes. Figure~\ref{figS3} shows the real and imaginary parts of the electric and magnetic dipole moments induced by an incident planewave (same as Fig.~1(c) in the main text) as calculated from exact fullwave calculations at real frequencies (empty markers) and as retrieved from the QNM approach using Eqs.~(\ref{eqS:ED-induced-QNM}) and (\ref{eqS:MD-induced-QNM}). One observes that Mode~\Romannum{1} (red solid lines) brings the major contribution on the induced dipole moments. Mode~\Romannum{2} (blue solid lines), thanks to the QNM distribution engineering done in the main text [Fig.~2], leads to a small yet non-negligible contribution on $\text{Re}\left[p_x (\omega) \right]$ that is almost constant in frequency. The exact results are recovered with good accuracy by summing over 13 QNMs (yellow dashed line). The agreement continues to improve by summing over more QNMs. Besides showing that the real frequency response of the plasmonic dolmen can be recovered with few QNMs, this reconstruction confirms once again the validity of our theoretical formalism for the multipole analysis of electromagnetic scattering with QNMs.

\section{Computation of the energy coupled by scattering to a dielectric waveguide}
\label{secS:nanowire}

Here, we describe how the power coupled to the dielectric nanowire via scattering by the plasmonic dolmen was computed (leading to Figs.~3(b)-(c)) in the main text. We consider a dielectric wire waveguide that is infinite along the $z$-direction and with a finite cross-section in the $xy$-plane. At frequency $\omega$, the waveguide supports forward and backward propagating (guided and leaky) modes, which will be denoted as $\Phi^{(q+,\omega)} = | \mathbf{E}^{q+},\mathbf{H}^{q+} > \exp \left[ i k_q(\omega) z \right]$ and $\Phi^{(q-,\omega)} = | \mathbf{E}^{q-},\mathbf{H}^{q-} > \exp \left[ - i k_q(\omega) z \right]$, respectively with $\mathbf{E}^{q\pm}$ and $\mathbf{H}^{q\pm}$ being the eigen electric and magnetic fields and $k_q(\omega)$ the Bloch wave vector of the forward-propagating mode. Depending on whether the mode is guided or leaky, $k_q(\omega)$ can either be real or complex.

When the waveguide is illuminated by an incident field, the total field in space be expressed as
\begin{equation}
\Phi^{(\omega)} = | \mathbf{E},\mathbf{H} > = \sum_{q=1}^N \alpha_{q+} \Phi^{(q+,\omega)} + \alpha_{q-} \Phi^{(q-,\omega)},
\end{equation}
with $\alpha_{q\pm}$ denoting the excitation coefficient of the mode $\Phi^{(q\pm,\omega)}$. From the Lorentz reciprocity theorem, we can reach the orthogonality relation~\cite{snyder2012optical}
\begin{equation}
\int \left( \mathbf{E}^{q-} \times \mathbf{H}^{p+} - \mathbf{E}^{p+} \times \mathbf{H}^{q-} \right) \cdot \mathbf{z} d\mathbf{x} d\mathbf{y} = F^{(p,\omega)} \delta_{pq},
\end{equation}
with $\delta$ the Kronecker delta and $F^{(p,\omega)}$ a complex constant. Numerically, the integral should be calculated over the cross-section of the entire computational domain, including in the medium surrounding the waveguide and the perfectly-matched layers.

The excitation coefficients are obtained using the orthogonality relation by
\begin{equation}\label{eqS:excitation}
\alpha_q = \frac{\int \left( \mathbf{E}^{q-} \times \mathbf{H} - \mathbf{E} \times \mathbf{H}^{q-} \right) \cdot \mathbf{z} d\mathbf{x} d\mathbf{y}}{F^{(q,\omega)}}.
\end{equation}
Knowing the excitation coefficient, the energy carried by the forward-propagating guided mode $\Phi^{(q+,\omega)}$ can be readily obtained via the Poynting vector as
\begin{equation}\label{eqS:power}
W^+ = \frac{1}{2} | \alpha_q|^2 \int \text{Re}\left[ \mathbf{E}^q \times \mathbf{H}^{q\star} \right] \cdot \mathbf{z} d\mathbf{x} d\mathbf{y}
\end{equation}
and similarly for the energy $W^-$ carried by backward-propagating mode $\Phi^{(q-,\omega)}$, leading to a total energy $W=W^+ + W^-$.

For the results of Fig.~3 of the main text and of Fig.~\ref{figS4} below, the energy coupled to the guided modes was calculated by substituting the total field $\mathbf{E}$ and $\mathbf{H}$ by the scattered field $\mathbf{E}^\text{s}$ and $\mathbf{H}^\text{s}$. The integrals in Eqs.~(\ref{eqS:excitation}) and (\ref{eqS:power}) were performed over $xy$ planes on either sides of the plasmonic resonator.

\section{Excitation of the Janus mode by an electric dipole and side dependent coupling}
\label{secS:dipole}

\begin{figure}[H]
   \centering
   \includegraphics[width=81.3636mm]{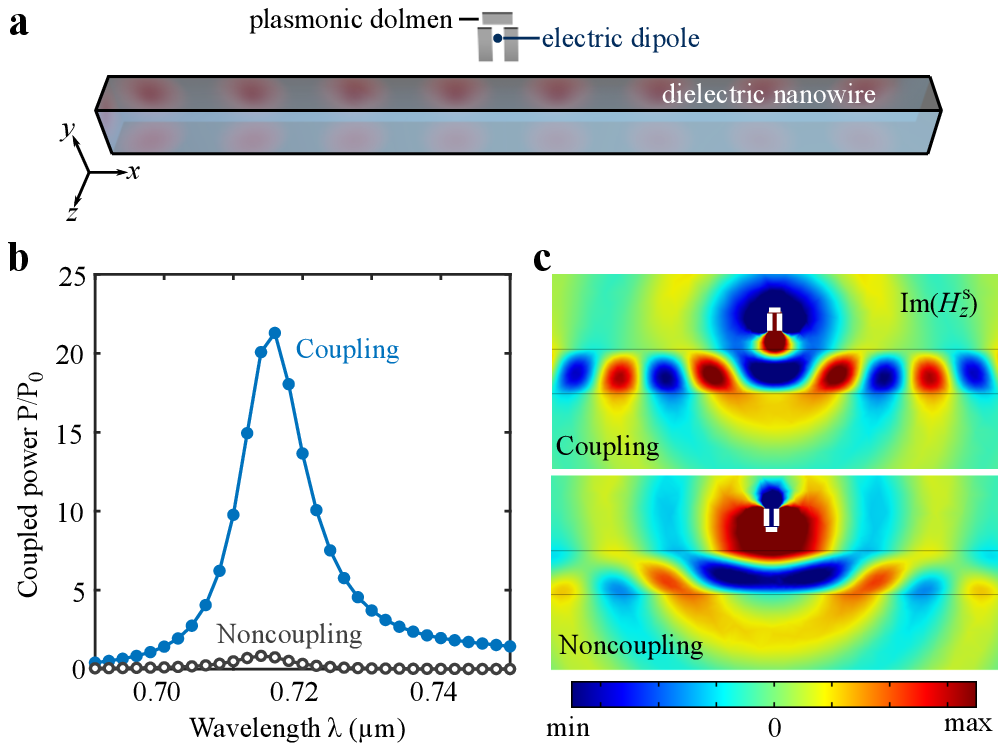}
   \caption{Side-dependent coupling to a nanowire waveguide via a Janus mode, when excited by an dipole source. \textbf{(a)} An electric dipole oriented along the $x$-direction is placed at the center of the plasmonic dolmen. \textbf{(b)} The energy carried by the guided modes as a function of wavelength and of the resonator orientation shows a strong side-dependent coupling to the nanowire mode. \textbf{(c)} Field maps of $\text{Im}[H_z^\text{s}]$ for the coupling and noncoupling situations at $\lambda=0.713$ $\mu$m.}
       \label{figS4}
\end{figure}

In the main text, we show that the plasmonic dolmen enables efficient coupling of a linearly-polarized planewave to the mode of a dielectric nanowire. Because the design was performed independently of the excitation, we expect that the same effect could be achieved with different sources. Here, we study the case where the nanowire-resonator system is excited by a dipole source. The geometric and material parameters of the system are the same as those used for Fig. 3 in the main text. An electric dipole oriented along the $x$-direction is placed at the center of the plasmonic dolmen, see Fig.~\ref{figS4}(a). In this way, we expect that Mode~\Romannum{3}, which radiates as an electric dipole along $y$, will not be excited. Figure~\ref{figS4}(b) shows the energy carried by the guided modes as a function of wavelength. Similarly to the planewave excitation, we find a strong side-dependent coupling to the waveguide mode with an efficiency ratio of about 23 at the resonance wavelength of 0.713 $\mu$m. The designed plasmonic dolmen therefore acts as a very efficient and controllable light coupler to waveguides.

$\quad$

\begin{thebibliography}{46}
\bibitem[{not({\natexlab{b}})}]{note1}
\bibinfo{note}{Complete convergence tests and comparisons will be provided in a
  future work.}
\end{thebibliography}
\end{document}